\documentclass{article}
\usepackage{spconf,amsmath,graphicx}
\usepackage{xcolor}
\usepackage{makecell,multirow,color,booktabs}
\usepackage{subcaption}
\usepackage{listings}
\usepackage{array}
\usepackage{tabularx}

\title{One-class knowledge distillation for spoofing speech detection}
\name{Jingze Lu$^{1,2}$, Yuxiang Zhang$^{1,2}$, Wenchao Wang$^{1}$, Zengqiang Shang$^{1}$, 
Pengyuan Zhang$^{1,2,*}$\thanks{*Corresponding author}\thanks{ 
This work is partially supported by the National Key Research 
and Development Program of China (No.~2021YFC33201033)}}

\address{$^1$Key Laboratory of Speech Acoustics and Content Understanding, Institute of Acoustics,\\
Chinese Academy of Sciences, Beijing, China\\
$^2$University of Chinese Academy of Sciences, Beijing, China\\
\small \tt \{lujingze, zhangyuxiang, wangwenchao, shangzengqiang, zhangpengyuan\}@hccl.ioa.ac.cn}
\begin{document}
%
\maketitle
\begin{abstract}

The detection of spoofing speech generated by 
unseen algorithms remains an unresolved challenge.
One reason 
for the lack of generalization ability is traditional detecting systems
follow the binary classification paradigm, 
which inherently assumes the possession of prior knowledge of spoofing speech.
One-class methods attempt to learn the distribution of 
bonafide speech and are inherently 
suited to the task where spoofing speech 
exhibits significant differences. 
However, training a one-class system using only bonafide speech is challenging.
In this paper, we introduce a teacher-student framework 
to provide guidance for the training of a one-class model.
The proposed one-class knowledge distillation method
outperforms other state-of-the-art methods on the~ASVspoof 21DF dataset and InTheWild dataset, 
which demonstrates its superior generalization ability.

\end{abstract}
\section{Introduction}
\label{sec:intro}

With the development of various text-to-speech (TTS) and 
voice conversion (VC) algorithms, 
a large amount of incredibly realistic synthesized speech 
could be generated at low cost.
In response to the potential threat 
posed by synthesized speech,
many detection systems have been 
developed by the research community.
The current mainstream detection scheme 
is a two-part paradigm, with a feature extractor front-end,
and a back-end for classification.
Researchers have explored 
different front-ends, such as STFT~\cite{zhang21da_interspeech}, CQCC~\cite{todisco2017constant}, 
and Wav2Vec~\cite{wang2021investigating,tak22_odyssey}, as well as different 
back-ends, such as RawNet~\cite{tak2021end} and AASIST~\cite{jung2022aasist}, 
for detecting fake utterances.

However, the lack of generalization ability 
remains an unsolved issue for current detection models. 
Some studies~\cite{asv_21_towards,muller_in_the_wild} 
indicate that anti-spoofing countermeasures (CMs) 
suffer a significant performance 
decline when facing unseen spoofing attacks, channel coding, 
compression coding, and other situations.
To improve the generalization ability,
some data augmentation methods have been proposed, including 
RawBoost~\cite{rawboost} and copy-synthesis method~\cite{vocoder_copy}.
In addition, \cite{wang2023investigating} introduces active learning (AL) 
to select useful training data to improve the generalization ability of the models.
In \cite{wang19i_interspeech}, domain adversarial methods are used to eliminate 
the differences between the target domain and the source domain.

In contrast to the above methods, 
in \cite{zhangyou_oneclass}, 
the failure of anti-spoofing CMs on unseen spoofing attacks
is attributed to the fact that they 
formulate the problem as a binary classification task.
The traditional binary classification training paradigm
inherently assumes that fake speech shares a similar distribution, 
which is not accurate.
It is unrealistic to assume that prior knowledge of the spoofing attack is
sufficient when potential attackers have a variety of synthesis methods.

\begin{figure*}[htb]
  \centering
  \includegraphics[width=0.98\linewidth]{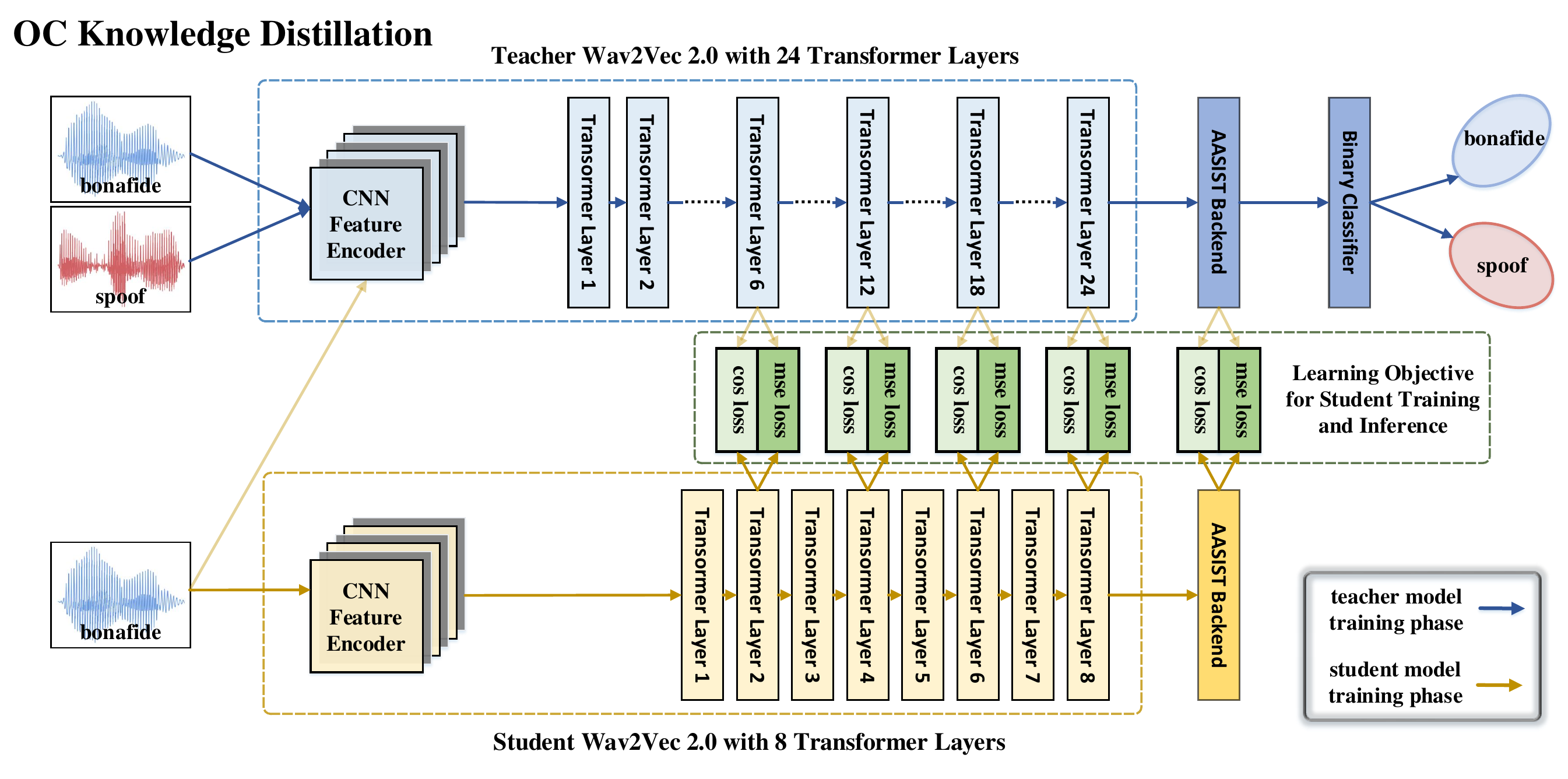}
  \caption{The pipeline of our proposed one-class knowledge distillation (OCKD)
  spoofing speech detection method. The teacher model contains a 24-layer Wav2vec 
  feature extractor and an AASIST backend.
  The student model has a similar structure to the teacher 
  model, but the Transformer layer of the feature extractor is 
  compressed to 8 layers.
  During the training phase of the student model, 
  the parameters of the teacher model are frozen.}
  \label{fig:kd}
\end{figure*}





Unlike binary classification, 
some studies~\cite{zhangyou_oneclass,oc_svm} utilize
one-class classification methods to improve the 
generalization ability of the anti-spoofing CMs.
One-class methods are inherently 
suited to tasks where spoofing speech 
exhibits significant differences.
The key point of one-class method
is to learn the distribution of real utterances 
and set an 
appropriate boundary around it.
Any speech outside the boundary could be considered fake.
Another potential benefit of one-class method
is that bonafide speech is 
much easier to collect compared to spoofing speech.
However, training a one-class spoofing speech detection 
system using only bonafide speech is challenging.
Speech is a coupling of multiple information.
For some information, such as semantics, 
bonafide and spoofing speech may share the same feature space.
Therefore, to obtain a distribution that adequately represents bonafide, 
we employ a teacher model with prior information to 
provide guidance for training the one-class model.

In this work, inspired by one-class classification methods and teacher-student frameworks, 
a one-class 
knowledge distillation approach is proposed
to improve the generalization ability of spoofing 
speech detection systems.
Different from previous knowledge distillation frameworks 
for model compression, 
the proposed framework 
conforms to the paradigm of anomaly sample detection.
In our framework, the teacher model
is a traditional binary classification model that
receives both bonafide and spoofing speech.
The student model only has access to bonafide speech
and focuses on learning the bonafide distribution from the teacher model.
The similarity of the output features of 
the teacher model and the student model is used 
for spoofing speech detection.
For real samples, the student has 
learned the 
representations 
from the teachers, resulting in a high similarity.
When faced with a variety of synthesized utterances, 
the student model is ignorant of them while the teacher 
model has prior information,
resulting in a low similarity.
Experimental results on ASVspoof 21DF and InTheWild datasets 
have demonstrated the generalization capability of our 
method against unknown attack algorithms.

In conclusion,
in response to the generalization problem of current 
anti-spoofing CMs, 
we propose a one-class 
knowledge distillation (OCKD) method for detecting spoofing speech. 
The proposed method outperforms other
state-of-the-art (SOTA) systems facing various unseen attacks 
by learning a distribution of bonafide speech.

\section{Method}
\label{sec:method}

Figure \ref{fig:kd} shows the pipeline of the proposed method, 
where the teacher model and the student model have a similar structure, 
including a Wav2Vec 2.0 front-end and an AASIST back-end.
In this section, we will introduce the motivation of 
designing the pipeline, and provide a detailed 
description of the modules and learning objectives.

\subsection{One-Class Knowledge Distillation 
Spoofing Speech Detection Method}
\label{ssec:ockd}
The key idea of one-class classification methods is to
learn a dense feature space of the target class distribution
and to distinguish samples far from this feature space as non-target samples.
Inspired by one-class methods, 
we intorduce a one-class knowledge distillation (OCKD) method 
to improve the generalization ability of speech anti-spoofing CMs.

The proposed OCKD method 
can be mainly divided into two parts, a teacher model 
and a student model.
The teacher model is trained with 
both bonafide and spoof speech, 
while the student model is trained 
using only bonafide speech.
With such a design, on the one hand, 
the teacher model can learn the 
differences between bonafide and spoof samples, 
and quickly learn a feature space
of bonafide samples.
The student model, on the other hand,
is not disturbed by the 
In-Distribution (ID) spoofing samples and 
focuses on 
learning the feature space of the bonafide speech.
The training of the teacher model and the student 
model are mutually independent.
When training the student model, 
the parameters of 
the teacher model are not updated.

For the teacher model, 
we follow the traditional 
structure of spoofing speech detection,
a feature extractor front-end and a 
classification back-end.
We adopt the same model architecture as in
\cite{tak22_odyssey}, with a Wav2Vec 2.0 front-end and 
an AASIST back-end.
The motivation for choosing such model structure 
as the teacher model is twofold.
Firstly, Wav2Vec 2.0 is a 
self-supervised-learning (SSL)-based feature
extractor.
During the training of Wav2Vec, 
only a large amount of 
bonafide speech is used, without the use of 
spoofing speech.
Therefore, fine-tuning the pre-trained Wav2Vec 
on the task of spoofing speech detection can 
achieve a reliable feature space of bonafide speech.
In addition, AASIST is an end-to-end, integrated 
spectrotemporal graph attention network.
The same structure used in \cite{tak22_odyssey} achieves 
state-of-the-art performance
on unseen attacks.
We believe that a teacher model
with greater generalization ability for 
unseen attacks is more helpful 
for the training of the student model.

The student model has a similar structure
as the teacher model, while 
the number of Transformer layers of 
Wav2Vec is compressed from 24 to 8.
There are two purposes for compressing the student model.
Firstly, compressing the model can 
accelerate the training of the student model.
In addition, the student model receives 
less data compared to the teacher model.
Compressing the model can prevent overfitting.

\subsection{Learning Objective}
\label{ssec:objective}
The learning objective of the student model is to
generate representations of bonafide speech similar to
the teacher model.
To achieve this objective, 
for a bonafide utterance, the embedding output 
by the student model should be 
close to that of the teacher model.
In the training process of the student model,
we construct the loss function 
whose target is the output embedding 
of the teacher model.
We use the mean square error (MSE) loss $\mathcal{L}_{mse}$
to measure the distance 
between embeddings.
However, the MSE loss is very sensitive
to outliers and is more difficult to 
train for models with a large number of parameters.
Therefore, we also introduce a loss function $\mathcal{L}_{cos}$
based on cosine similarity.
For a training set $\mathcal{B}$ of all bonafide speech,
the total loss function could be expressed as,
\begin{equation}
  \mathcal{L}_{total}=\mathcal{L}_{cos}+\lambda \mathcal{L}_{mse}
\end{equation}
where $\lambda$ is a constant value, which should be 
set according to the loss value of $\mathcal{L}_{cos}$ and $\mathcal{L}_{mse}$ 
to ensure they
are in the same order of magnitude.
$\mathcal{L}_{mse}$ can be expressed as 
$\mathcal{L}_{mse}=\frac{1}{N}\sum_{b_i \in \mathcal{B}}(T(b_i)-S(b_i))^2$.
$\mathcal{L}_{cos}$ can be expressed as 
$\mathcal{L}_{cos}=\frac{1}{N}\sum_{b_i \in \mathcal{B}}(1-\frac{\langle T(b_i), S(b_i)\rangle}{\Vert T(b_i)\Vert_2\Vert S(b_i)\Vert_2})$,
where~$\langle\cdot\rangle$~donates the inner product, and 
$\Vert\cdot\Vert_2$ donates the computation of the 2-Norm.



\cite{lee22q_interspeech} indicates that for the task of 
spoofing speech detection, 
different Transformer layers of Wav2Vec 
play different roles.
Therefore, the student model should learn 
from the teacher model at different levels.
The student model of the proposed 
OCKD has 8 Transformer layers, 
which is only one-third of the teacher model.
We use the hidden embedding $\{s_2, s_4, s_6, s_8\}$
to learn from $\{t_6, t_{12}, t_{18}, t_{24}\}$, 
where $t_i$ and $s_i$ are the 
output of the $i$th Transformer layer 
of teacher and student, respectively.
The back-end of the teacher model is 
also important for spoofing speech detection, 
so $t_A$ is also set to be the learning target of $s_A$, 
where $t_A$ and $s_A$ are the hidden embedding 
output by AASTST back-ends of teacher and student model, respectively.




Both the teacher model and 
the student model are used to
develope the inference model.
During the testing phase, 
for an unknown utterance $x$,
similarity
between $\{s_2, s_4, s_6, s_8, s_A\}$
and $\{t_6, t_{12}, t_{18}, t_{24}, t_A\}$ is used for inference.
For bonafide utterances, the student model
has learned the representations from the teachers,
resulting in a high similarity in the embeddings.
When confronted with various unseen attack algorithms,
the student model is ignorant of them.
In contrast, the teacher model possesses prior knowledge 
of the spoofing speech, 
leading to a low similarity in the embeddings.





\begin{table*}[tb]
  \caption{EERs (\%) $\downarrow$ of teacher model and student model 
  of proposed OCKD method on differentd datasets.
  student\_mse and student\_cos donate the student model 
  with single learning objective, student\_total use both learning objective.}
  \setlength\tabcolsep{3pt}
  \label{tab_result_1}
  \centering
  \begin{tabularx}{\textwidth}{cc|*{6}{X<{\centering}}|X<{\centering}cccccccccccccccccc}
    \toprule[1.2pt]

    \textbf{architecture}&\textbf{model}&\small \tt{19LA}&\small \tt{21LA}&\small \tt{21DF}&\small \tt{21LA\_hid}&
    \small \tt{21DF\_hid}&\small \tt{InTheWild}&\small \tt{Pooled}\\
    \midrule
    \multirow{4}{*}{\makecell[c]{\textbf{Wav2Vec}\\ \textbf{+AASIST}}}&\textbf{teacher}
    &\textbf{0.22}&\textbf{0.82}&2.85&15.29&12.02&10.48&6.36\\
    &\textbf{student\_mse}&1.44&1.81&11.90&27.50&25.40&26.49&18.46\\
    &\textbf{student\_cos}&0.35&0.91&\textbf{2.27}&\textbf{14.25}&\textbf{9.34}&8.38&5.96\\
    &\textbf{student\_total}&0.39&0.90&\textbf{2.27}&14.65&9.54&\textbf{7.68}&\textbf{5.88}\\
  \bottomrule[1.2pt]
\end{tabularx}
\end{table*}


\section{Experiments and Results}
\label{sec:result}

\subsection{Datasets and Metrics}
\label{ssec:dataset}

To investigate the generalization ability of the proposed method,
experiments are conducted on several different datasets.
For all models, we utilize the training set of the ASVspoof 2019 LA (19LA)~\cite{wang2020asvspoof} for training, 
which is an influential dataset in spoofing speech detection.
For the teacher model, all 25380 samples in 19LA training set
are used.
While for student model, only the bonafide samples (num 2580)
are used for training.
The test sets include the evaluation sets of 19LA, 
ASVspoof 2021 LA (21LA) and ASVspoof 2021 DeepFake (21DF)~\cite{asv_21_towards}.
Utterances of the 21LA dataset are transmitted over various channels. 
The 21DF dataset 
collects about 600K utterances processed with 
various lossy codecs typically used for media storage, 
which is an influential dataset for generalization validation.
21LA and 21DF have a hidden track, 
in which the non-speech segments are trimmed.
In addition, the proposed method
is validated on the InTheWild dataset~\cite{muller_in_the_wild}, 
which is a more challenging dataset
whose data is collected from the real world.
The equal error rate (EER) is used as the evaluation metric,
which is defined as the point where the 
false acceptance rate (FAR) and the false
rejection rate (FRR) are equal.





\subsection{Model Architecture and Details of Systems Implementation}
\label{ssec:detail}
For the teacher model, we adopt the same 
structure as~\cite{tak22_odyssey}, with a Wav2Vec 2.0 front-end 
and an AASIST back-end.
The pre-trained model used
is Wav2Vec2-xlsr. 
During the training process of the teacher model, 
the pre-trained wav2vec 2.0 
model is optimized jointly with the AASIST backend.
The student model has a similar 
structure as the teacher, while 
the number of Transformer layers 
is compressed to 8.
During the training process of the student model, 
parameters of the teacher model are frozen.
All models are trained using Adam optimizer with $\beta_1 = 0.9$, $\beta_2=0.98$,
$\epsilon=10^{-8}$ and weigth decay $10^{-4}$. 
The teacher model utilizes a 
CrossEntropy loss function with weight of $\{0.1,0.9\}$ to balance the training set.
The learning objective of the student 
model is expressed in section \ref{ssec:objective}, 
where $\lambda$ is set to a constant value $10^{-5}$. 
The learning rate is fixed at $10^{-6}$. 
Training is conducted over 100 epochs and a batchsize of 32.
Rawboost~\cite{rawboost} is used as data augmentation method.


\subsection{Reuslts and Analysis}
\label{ssec:result}

Tabel \ref{tab_result_1} shows the comparison of the teacher model 
and the student model of proposed OCKD method on different datasets.
For 19LA and 21LA datasets, the EER of the student has slightly decreased.
This may be because the eval sets of 19LA and 21LA 
contain many ID samples as the training set.
For samples that have similar distribution as the training set, 
binary classification 
method yields superior results.
On other eval sets, the student model achieves lower EERs.
Among them, the 21DF and the InTheWild 
dataset are two commonly used datasets 
for evaluating the generalization ability.
21DF dataset contains
more than 100 different spoofing attack algorithms.
InTheWild dataset is collected from the real world.
Obtaining performance gains 
on these two datasets 
demonstrate that the proposed 
one-class knowledge distillation 
method enables the student 
model to learn the distribution of real speech and 
effectively improving the generalization ability on unseen attacks.
21LA and 21DF hidden sets are the official subsets of 
ASVspoof 2021 dataset.
These two subsets trim the non-speech segments, 
which can lead to performance degradation of anti-spoofing CMs~\cite{10224301}.
The reason for such performance degradation is that
CMs may overfit to the length
of non-speech segments.
The propsed OCKD method also 
obtains performance improvements on both datasets.
For all datasets, 
the overall pooled EER of the student model
decreases from 6.36\% to 5.88\%.

Tabel \ref{tab_result_1} also shows the ablation results of the
learning objective.
For the two learning objectives introduced in this work, 
the performance of the MSE loss is poor.
This could be attributed to the fact that the hidden embeddings 
used to be the target of learning objective have large dimension.
It is hard to enforce the embeddings of the student model to 
align with those of the teacher model.
In contrast, the cosine similarity loss achieves good results.
Using two loss functions simultaneously
achieves slight improvements compared to using only cosine loss.

\begin{table}[htb]
  \caption{Comparison with recently proposed state-of-the-art systems 
  on the ASVspoof 2021 DF dataset and InTheWild dataset.
  EER(\%) $\downarrow$ is used as metric.}
  \label{tab:dfresults}
  \centering
  \begin{tabular}{ccc}
    \toprule[1.2pt]
    \textbf{Eval Set}&\textbf{Model} &\textbf{EER (\%) $\downarrow$}\\
    \midrule
    \multirow{5}{*}{\small \tt{2021 DF}}
    &ResNet\cite{chen21b_asvspoof}&16.05\\
    &Hubert+LCNN\cite{wang22_odyssey}&12.39\\
    &Wav2Vec XLS-128\cite{9747768} &4.98 \\
    &Wav2Vec+LCNN\cite{wang22_odyssey}&4.75\\
    &Wav2Vec+Linear\cite{10094779}&3.65 \\
    &Wav2Vec+AASIST\cite{tak22_odyssey}&2.85\\
    &\textbf{OCKD(proposed)}&\textbf{2.27}\\
    \midrule
    \multirow{4}{*}{\small \tt{InTheWild}}&RawNet2\cite{muller_in_the_wild}&33.94 \\
    &CQT+CVNN\cite{muller23_interspeech}&26.95\\
    &Wav2Vec+Linear\cite{10094779}&16.17 \\
    &\textbf{OCKD(proposed)}&\textbf{7.68}\\
  \bottomrule[1.2pt]
  
  
\end{tabular}
\end{table}

Tabel \ref{tab:dfresults} 
shows the comparison of the results of our proposed 
OCKD model with other recently 
proposed SOTA
systems on 21DF and InTheWild dataset.
Our system 
outperforms other SOTA
systems on these two datasets,
which proves that our system generalizes 
well to large amounts of unseen attack algorithms and 
real world scenarios.

\section{Conclusion}
\label{sec:conclusion}

This paper focuses on improving the generalization ability of spoofing 
speech detection systems facing unseen attack samples.
We attribute the lack of generalizability 
of traditional anti-spoofing systems 
to their binary classification properties and introduce a one-class 
method.
To provide guidance for the training of one-class model,
we adopt a teacher-student framework.
In the proposed one-class knowledge distillation method, 
teacher model
is trained with bonafide and spoof utterances, 
and learns the distribution of bonafide speech. 
Then the student model, which only has access to bonafide speech,
learns this distribution from the teacher model.
The similarity between the hidden embeddings of teacher 
and student is used for detection.
The proposed method outperforms other SOTA systems
on 21DF and InTheWild datasets, 
which demonstrates its generalization ability 
for various unseen spoofing attacks and scenarios.



\vfill
\pagebreak


\bibliographystyle{IEEEbib}
\bibliography{strings,refs}

\end{document}